\documentstyle[epsfig]{aipproc}

\def\be{\begin{equation}}
\def\ee{\end{equation}}
\def\ba{\begin{eqnarray}}
\def\ea{\end{eqnarray}}

\def\ga{\mathrel{\mathpalette\fun >}}
\def\fun#1#2{\lower3.6pt\vbox{\baselineskip0pt\lineskip.9pt
        \ialign{$\mathsurround=0pt#1\hfill##\hfil$\crcr#2\crcr\sim\crcr}}}

\begin{document}
\title{Model-Independent Measurement of the Primordial Power Spectrum}

\author{Yun Wang, David N. Spergel, \& Michael A. Strauss}
\address{Department of Astrophysical Sciences \\
Princeton University, Princeton, NJ 08544}


\maketitle

\begin{abstract}

In inflationary models with minimal amount of gravity waves,
the primordial power spectrum of 
density fluctuations, $A_S^2(k)$, 
together with the basic cosmological parameters, completely specify
the predictions for the cosmic microwave background (CMB) anisotropy
and large scale structure.
Here we show how we can strongly constrain both $A_S^2(k)$ and the 
cosmological parameters by combining the data from the Microwave 
Anisotropy Probe (MAP) and the galaxy redshift survey from the Sloan 
Digital Sky Survey (SDSS). We allow $A_S^2(k)$ to be a free function, 
and thus probe features in the primordial power spectrum on all scales.
MAP and SDSS have scale-dependent measurement errors that decrease
in opposite directions on astrophysically interesting scales; they
complement each other and allow the measurement of the primordial power 
spectrum independent of inflationary models, giving us valuable
information on physics in the early Universe, and providing clues to
the correct inflationary model.

\end{abstract}


At present, inflation provides the simplest and most compelling solution
to the problems of the standard cosmology: the horizon problem,
the flatness problem, and the structure formation problem.
A generic prediction of inflationary models is the existence of 
primordial adiabatic Gaussian random-phase density fluctuations. 
The properties of these fluctuations are completely
determined by their power spectrum $A_S^2(k)$.
There currently exists a broad range of inflationary models; they make
different predictions for the primordial power spectrum $A_S^2(k)$.

It is customary to parametrize $A_S^2(k)$ as a featureless power law,
$A_S^2(k) \propto k^{n_S-1}$. However, there
are models for which the power-law index would change with scale \cite{Wang94}.
There are other models which predict primordial power
spectra with broken scale invariance,
which can not be parametrized by a featureless power law \cite{broken}.
There are physical reasons as well as tentative observational evidence 
\cite{Peacock}
which point to the possibility that the primordial power spectrum has breaks
in its power-law form. 
Given our ignorance of the true form of $A_S^2(k)$,
we feel that one should allow $A_S^2(k)$ to be a free function.
In this paper we show how we can measure $A_S^2(k)$ without assuming
that it is given by any specific model \cite{Wang99}.

We parametrize $A_S^2(k)$ as a step function,
with amplitudes $a_i$ ($i=1,2,...,20$) in log$\,k$ bins
($k_1=0.001\,h$Mpc$^{-1}$, and $k_2$-$k_{20}$ are equally
spaced in log$\,k$ from $k_1$ to $k=0.5h\,$Mpc$^{-1}$).
The $a_i$ ($i=1,2,...,20$) are taken to be independent variables.
For definiteness, we adopt a model in which the true primordial power
spectrum is $A_S^2(k)=1$ (the scale-invariant Harrison-Zel'dovich spectrum), 
and the cosmological parameters are $h=0.5$ (dimensionless Hubble 
constant), $\Omega_m=1$ (matter density in units of the critical density), 
$\Lambda=0$ (cosmological constant), $\Omega_b=0.05$ (baryon density in 
units of the critical density), $\tau_{ri}=0.05$ (reionization optical depth), and $b=0.83$ (galaxy bias parameter). 

The upcoming data from the Microwave Anisotropy Probe (MAP) {\cite{MAP} and
the Sloan Digital Sky Survey (SDSS) \cite{SDSS}
provide a unique opportunity for constraining inflationary models.
It is useful to expand the temperature fluctuations in the CMB
into spherical harmonics: 
$\delta T/T (\hat{\bf r})= \sum_{l,m} a_{T,lm} Y_{lm}(\hat{\bf r})$,
where $\hat{\bf r}$ is the unit direction vector in the sky, and
\be
\label{C_l def}
C_{Tl} \equiv \langle |a_{T,lm}|^2 \rangle= (4\pi)^2 \int \frac{dk}{k}\,
A_S^2(k)\, \left| \Delta_{Tl} (k, \tau=\tau_0) \right|^2.
\ee
$\Delta_{Tl} (k, \tau=\tau_0)$ is an integral over 
conformal time $\tau$ of the sources that generate the CMB fluctuations,
and $\tau_0$ is the conformal time today.

The power spectrum of mass fluctuations in the linear regime today is
\be
P(k)=P_0 \,k A_S^2(k)\, T^2(k),
\ee
where $P_0$ is a normalization constant, and $T(k)$ is the transfer 
function, which depends on physics at matter-radiation equality
and decoupling. The galaxy redshift survey from the SDSS
will allow a determination of $P_G(k)$,
the galaxy power spectrum in redshift space; it differs from $P(k)$
due to the bias between mass and galaxy density fields
and redshift distortion effects induced by peculiar velocities. We write
$P_G(k) = b^2 \left(1 +  {2 \over 3} \beta + {1 \over 5} \beta^2\right)
P(k) \equiv b_{\it eff}^2 P(k)$,
where $\beta \equiv \Omega_m^{0.6}/b$.
We take nonlinear effects into consideration by using analytical
formulae \cite{Wang99}. We assume the bias $b$ to be constant,
which is a good approximation on large scales \cite{bias}.
On small scales, bias must be scale-dependent and nonlinear.
Different galaxy samples (spirals, ellipticals, etc) can be used to test for
more complicated bias laws.

Both $C_{Tl}$ and $P_G(k)$ are proportional to the amplitude of $A_S^2(k)$.
Fig. 1 shows the accuracy in the determination of the bin amplitudes
of $k\,A_S^2(k)$; the 1-$\sigma$ error bars
are shown for using (a) MAP temperature data only, and
(b) SDSS data only. The errors have been estimated using the Fisher matrix
\cite{Wang99}.
Because of the finite resolution of the MAP satellite, the errors on
the $C_{Tl}$ increase rapidly with wavenumber for $k\ga 0.2\,h\,$Mpc$^{-1}$
(see Fig.1a), while the errors on $P_G(k)$ from the SDSS decrease with 
wavenumber (see Fig.1b), because on small scales one can average
over many $k$-modes.
MAP and SDSS thus quite naturally complement each other in
the determination of $A_S^2(k)$.
By combining MAP temperature and SDSS data,
the primordial power spectrum can be determined to around 16\% 
accuracy for $k\sim 0.01\,h\,$Mpc$^{-1}$, and to
around 1\% accuracy for $k\sim 0.1\, h\,$Mpc$^{-1}$,
assuming that the cosmological parameters are known \cite{Wang99}.
\begin{figure} 
\psfig{file=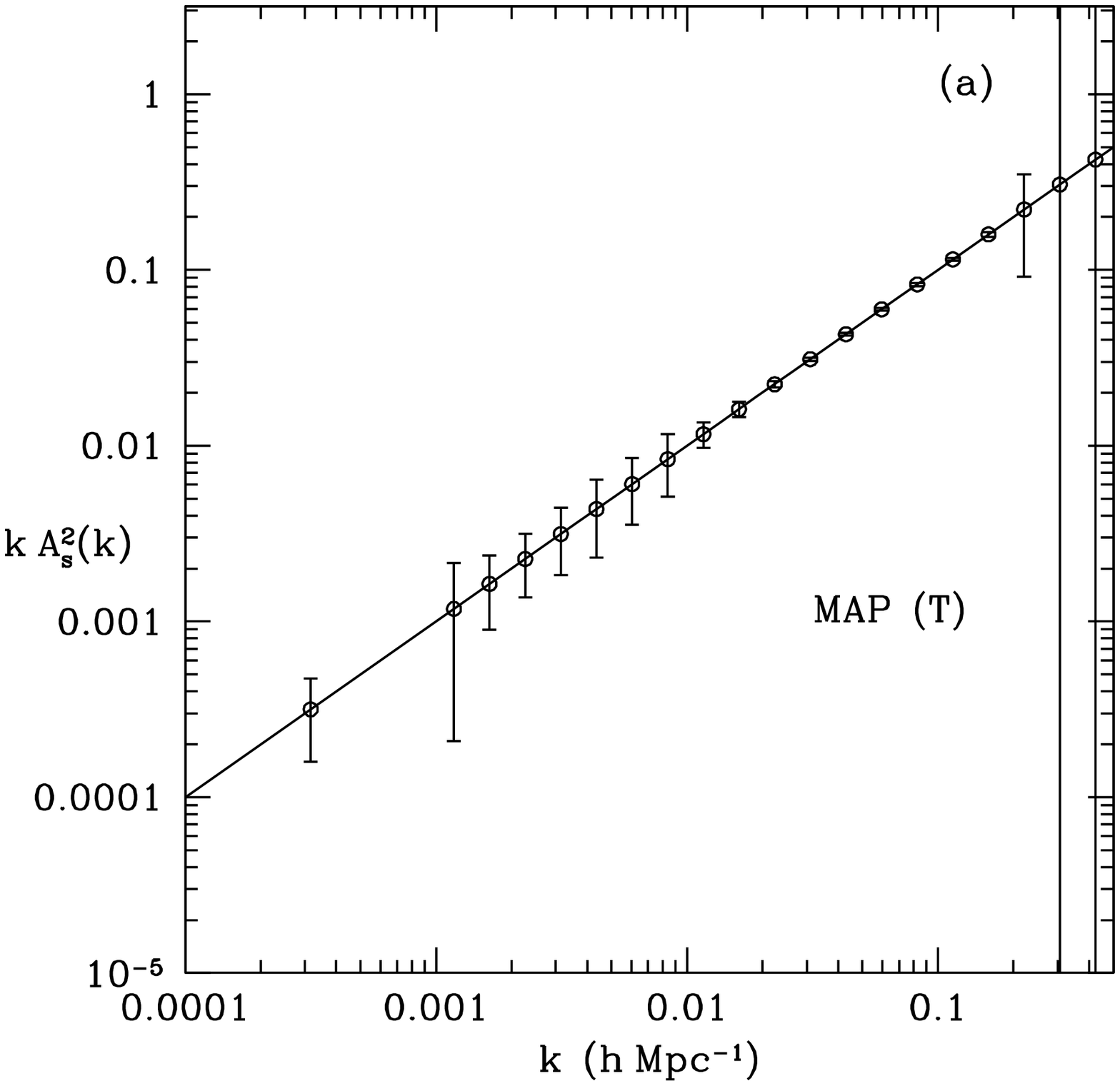,height=2.8in,width=2.8in}\hspace*{0.1in}
\psfig{file=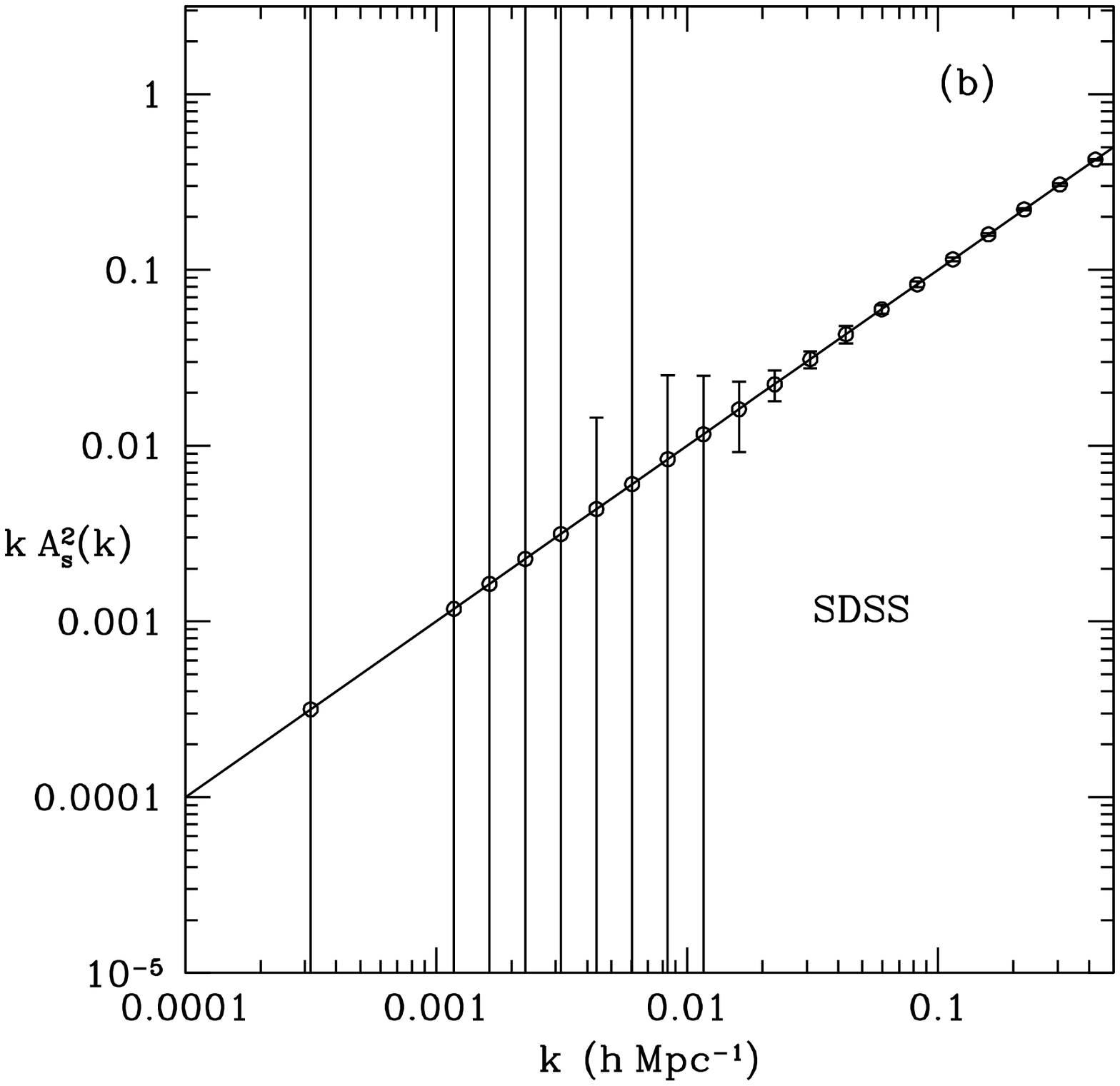,height=2.8in,width=2.8in}
\caption{The primordial power spectrum with 1-$\sigma$ 
error bars (assuming that the cosmological parameters are known)
for (a) MAP temperature data only; (b) SDSS data only.}
\label{fig:MAPSDSS}
\end{figure}

The determination of the primordial power spectrum
bin amplitudes $a_i$ does depend on the a priori knowledge of the
cosmological parameters.  We now estimate four 
parameters $h$, $\Omega_{\Lambda}$, $\Omega_b$, and $\tau_{ri}$,
in addition to the $a_i$.
The $a_i$ and the cosmological parameters cannot be simultaneously determined
using MAP temperature data only, because $\tau_{ri}$ is degenerate
with the $a_i$ for $k\ga 0.01\,h\,$Mpc$^{-1}$.
However, polarization data removes this degeneracy.
Figure 2(a) shows the resulting error bars on the 
primordial power spectrum if both the 
temperature and polarization data from MAP are used; they are 
significantly larger than in Figure 1(a) where only temperature data
is used. 

Our ability to use the MAP polarization data may be limited by our ability to
remove foregrounds. It is therefore important to have an alternative
to the MAP polarization data in parameter estimation.
Figure 2(b) shows the effect of combining the MAP temperature data with
the SDSS data in the
analysis, adding the galaxy bias $b_{\it eff}$ as a fifth parameter. 
The uncertainty in the determination of the $a_i$ increases by a factor 
up to 3 on small scales 
relative to the
case in which the cosmological parameters are known {\it a priori}. 
This difference only enters through error bars that are only of the 
order of a few percent.
The errors in the $a_i$ are not significantly increased by the inclusion
of cosmological parameters in the parameter estimation,
because most of the information on the cosmological parameters
comes from small angular scales, where the power spectrum can be determined
to great accuracy from the SDSS data.
Comparison of Figure 2(a) and Figure 2(b) illustrates the importance
of combining the MAP data with the SDSS data in 
the determination of the primordial power spectrum bin amplitudes $a_i$ \cite{Wang99}.
\begin{figure} 
\psfig{file=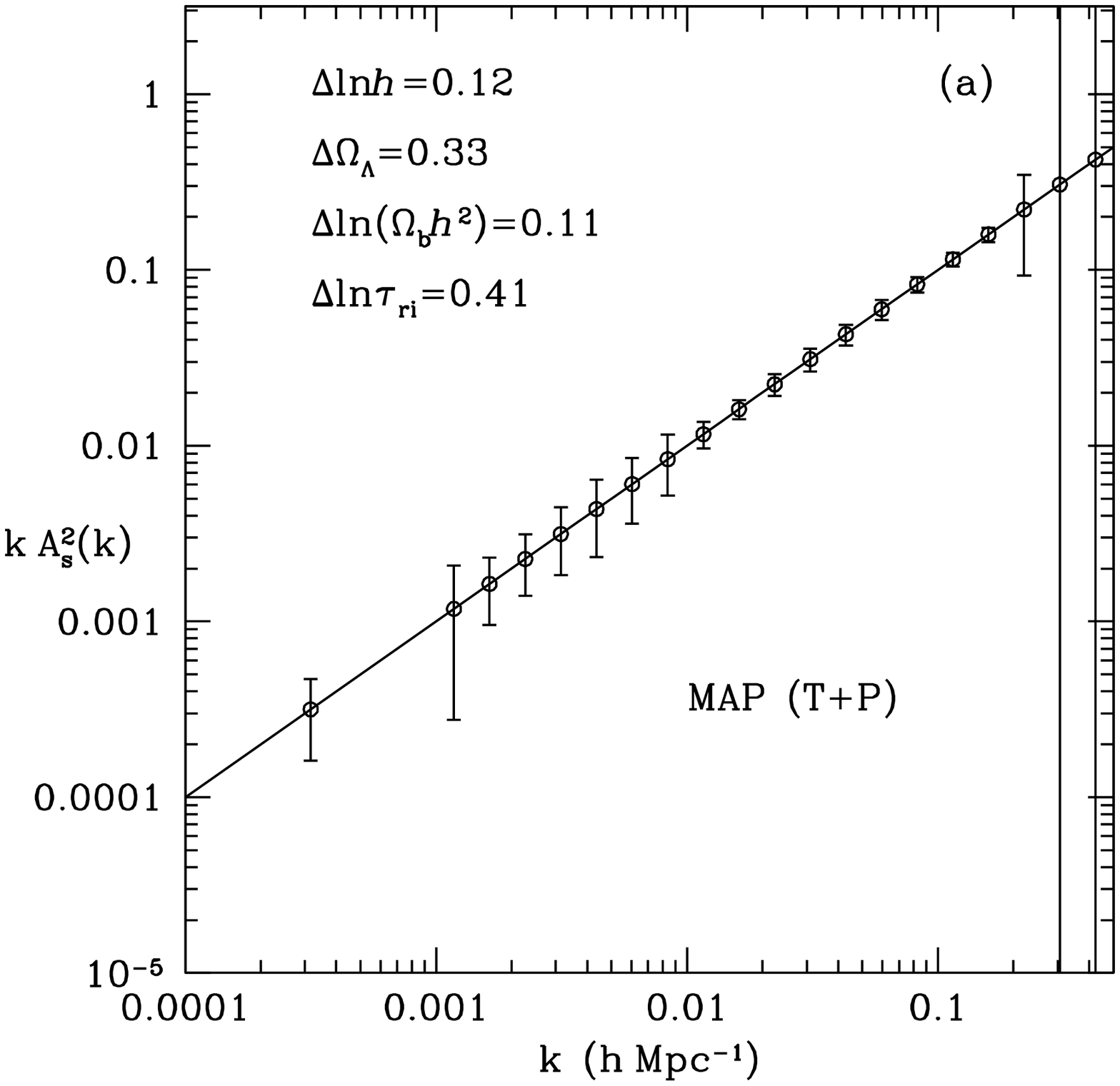,height=2.8in,width=2.8in}\hspace*{0.1in}
\psfig{file=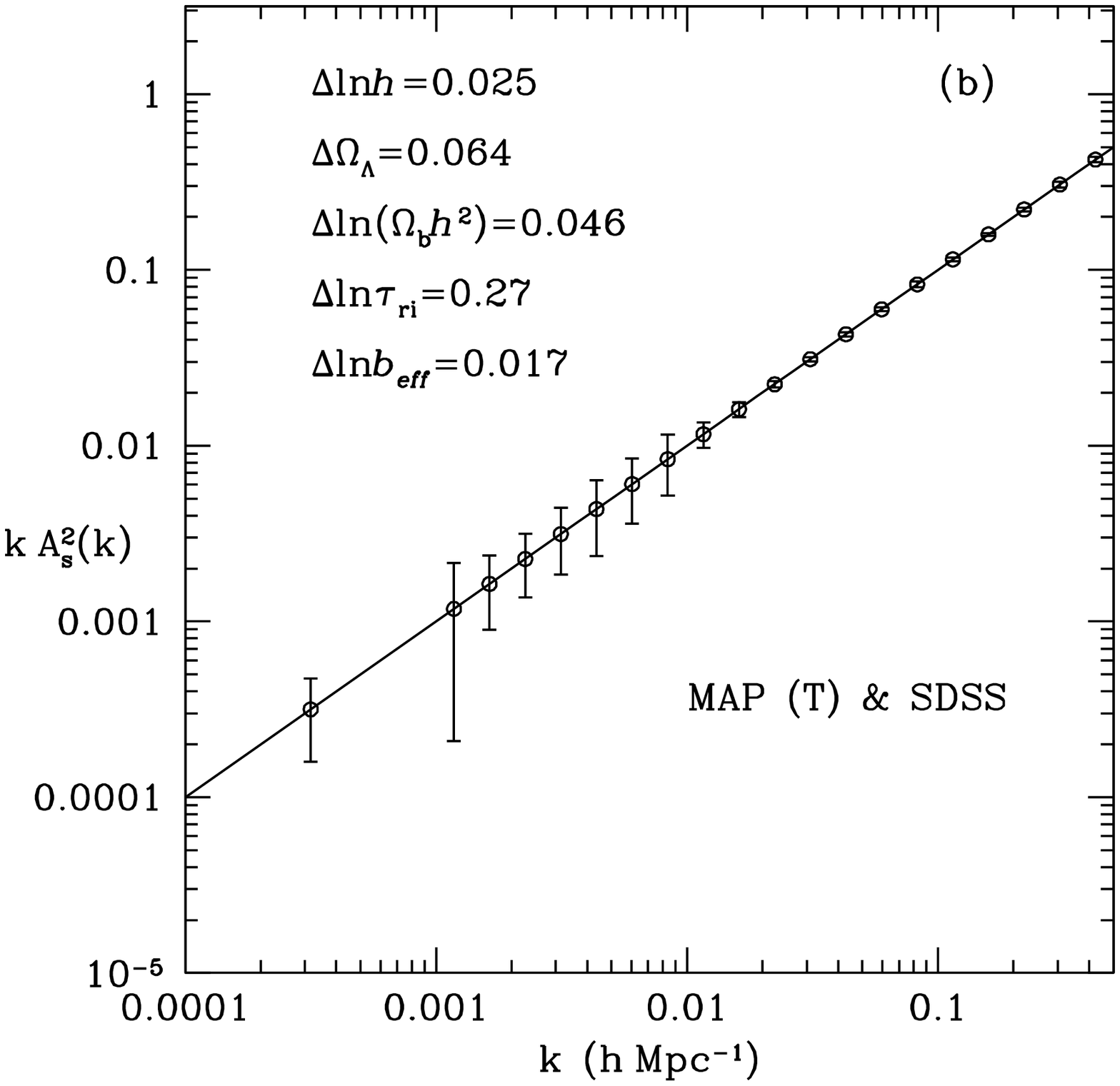,height=2.8in,width=2.8in}
\caption{The primordial power spectrum with 1-$\sigma$ error bars.
(a) Combined MAP temperature and polarization data,  
with four cosmological parameters 
($h$, $\Omega_{\Lambda}$, $\Omega_b$, $\tau_{ri}$)
included in the parameter estimation;
(b) Combined MAP temperature and SDSS data, with
five cosmological parameters
($h$, $\Omega_{\Lambda}$, $\Omega_b$, $\tau_{ri}$, $b$) 
included in the parameter estimation.}
\label{fig:MAPSDSSc}
\end{figure}

Assuming that inflation occurred, all the obtainable information
about the inflationary model is contained in 
the primordial power spectrum $A_S^2(k)$. 
From the combination of the MAP and SDSS data, we can obtain
a measurement of $A_S^2(k)$ without reference to
any specific inflationary model, assuming that the primordial
fluctuations are adiabatic. 
If the values of the cosmological parameters that best fit the data
are close to what we expect, but the primordial power spectrum differs
significantly from the predictions of the current inflationary models,
then it will be an indication of new physics in the early universe,
and it will provide a solid starting point for building new 
inflationary models. 

This work is supported in part by NSF grant AST98-02802, the MAP/MIDEX project, 
and NSF grant AST96-16901.

\end{document}